\let\latexdocument\document
\let\latexarabic\arabic

\documentclass{article}
\usepackage[margin=1.5in]{geometry}

\let\document\latexdocument
\let\arabic\latexarabic

\usepackage{times}
\usepackage{bm}
\usepackage{natbib}

\usepackage{amsmath} 
\usepackage{hyphenat}
\usepackage{bbold}
\usepackage{bbm}
\usepackage[T1]{fontenc}
\usepackage{amsmath}
\usepackage{amssymb}
\usepackage{prodint}
\usepackage{color}

\newtheorem{assumption}{Assumption}
\newtheorem{remark}{Remark}
\newtheorem{definition}{Definition}

\usepackage{mathtools}
\usepackage[usestackEOL]{stackengine}

\usepackage{accents}

\newcommand{\cadlag}{c\`{a}dl\`{a}g }

\usepackage{amsmath}
\usepackage{mathrsfs}
\newcommand{\EE}{\mathbb{E}}

\newcommand{\eps}{\varepsilon}

\newcommand{\F}{\mathcal{F}}

\newcommand{\vv}{v}

\newcommand{\R}{\mathbb{R}}

\newcommand{\1}{\mathbb{1}}

\newcommand{\gridsize}{h}%m}

\newcommand{\tmax}{\ensuremath{\tau}}%{\tau}

\newcommand{\argmin}{\text{argmin}}
\newcommand{\cupdot}{\mathbin{\mathaccent\cdot\cup}}
\newcommand\independent{\protect\mathpalette{\protect\independenT}{\perp}}\def\independenT#1#2{\mathrel{\rlap{$#1#2$}\mkern2mu{#1#2}}}

\usepackage[plain,noend]{algorithm2e}

\makeatletter
\renewcommand{\algocf@captiontext}[2]{#1\algocf@typo. \AlCapFnt{}#2} % text of caption
\def\@algocf@capt@plain{top}
\renewcommand{\algocf@makecaption}[2]{%
  \addtolength{\hsize}{\algomargin}%
  \sbox\@tempboxa{\algocf@captiontext{#1}{#2}}%
  \ifdim\wd\@tempboxa >\hsize%     % if caption is longer than a line
    \hskip .5\algomargin%
    \parbox[t]{\hsize}{\algocf@captiontext{#1}{#2}}% then caption is not centered
  \else%
    \global\@minipagefalse%
    \hbox to\hsize{\box\@tempboxa}% else caption is centered
  \fi%
  \addtolength{\hsize}{-\algomargin}%
}
\makeatother

\begin{document}

\title{ Estimation of time-specific intervention effects on
  continuously distributed time-to-event outcomes by targeted maximum
  likelihood estimation}
\author{\small Helene C. W. Rytgaard$^{1,*}$, 
F. Eriksson$^{1}$, and Mark J. van der Laan$^{2}$ \\
\small $^{1}$Section of Biostatistics, University of Copenhagen, Denmark\\ \small$^{2}$Devision of Biostatistics, University  of California, Berkeley}

\maketitle

\begin{abstract}

  Targeted maximum likelihood estimation is a general methodology
  combining flexible ensemble learning and semiparametric efficiency
  theory in a two-step procedure for estimation of causal
  parameters. Proposed targeted maximum likelihood procedures for
  survival and competing risks analysis have so far focused on events
  taken values in discrete time.  We here present a targeted maximum
  likelihood estimation procedure for event times that take values in
  \(\R_+\). We focus on the estimation of intervention-specific mean
  outcomes with stochastic interventions on a time-fixed treatment.
  For data-adaptive estimation of nuisance parameters, we propose a
  new flexible highly adaptive lasso estimation method for
  continuous-time intensities that can be implemented with
  \(L_1\)-penalized Poisson regression.  In a simulation study the
  targeted maximum likelihood estimator based on the highly adaptive
  lasso estimator proves to be unbiased and achieve proper coverage in
  agreement with the asymptotic theory and further displays efficiency
  improvements relative to a Kaplan-Meier approach.
 
\end{abstract}

\textbf{Keywords:} Targeted maximum likelihood estimation, survival
analysis, treatment effects, semiparametric model, efficient
estimation, causal inference

\section{Introduction}
\label{sec:introduction}

In recent years, semiparametric efficient and doubly robust estimators
\citep{vanRobins2003unified,tsiatis2007semiparametric} have gained
great popularity in many fields and are increasingly used to draw
inference about the effect of a treatment
\citep{robins1986new,hernanrobins}.  In survival analysis, such
methods provide an alternative to the widely used Cox regression model
\citep{cox1972regression} that relies on the assumption of
proportional hazards and requires correct specification of main
effects and interaction effects of treatment and confounders.  We
consider a standard survival analysis setting where a population of
subjects are followed over a period of time until an event of interest
occurs \citep{andersen2012statistical}. We suppose that covariates are
measured and a treatment decision, either randomized or conditional on
covariates, is made at the beginning of the follow-up period, and that
we are interested in the effect of the treatment decision on the time
until the event of interest happens.  The data are further
characterized by right-censoring, saying that a subject is
right-censored if the event of interest did not occur within the
subject-specific follow-up period.  In this work we
target % of estimation should be defined as
parameters of the g-computation formula that have a clear causal
interpretation under a set of structural assumptions
\citep{robins1986new,gill2001causal,vanRobins2003unified,hernanrobins}
and consider the general case with stochastic interventions
\citep[][Sections 6 and
7]{robins2004effects,dawid2010identifying,gill2001causal} on the
treatment decision on time-to-event outcomes; an important special
case is the average treatment effect on the \(\tmax\)-year risk of
death.

Our overall goal for estimation is to impose as few assumptions as
possible on the data-generating mechanism while providing efficient
and double robust estimation by optimizing the estimation procedure
for the target parameter specifically.  Targeted maximum likelihood
estimation \citep{van2006targeted,van2011targeted,van2018targeted}
provides a general methodology that combines cross-validated machine
learning \citep{van2006oracle,van2006cross,van2007super} and
semiparametric efficiency theory \citep{bickel1993efficient} for
constructing asymptotically efficient estimators for low-dimensional
parameters in infinite-dimensional models.  Targeted maximum
likelihood methods have been extensively developed for treatment
effect estimation in survival analysis settings where events are
observed on a discrete time-scale
\citep{moore2009application,moore2009covariate,moore2009increasing,
  stitelman2011targeted2,stitelman2011targeted,van2012targeted,benkeser2018improved,cai2018one},
but requires artificial discretization to be applied to events
observed in continuous time and may lead to instability and high
memory usage \citep{sofrygin2019targeted}. We propose a generalization
of existing targeted maximum likelihood methods to continuously
measured time-to-event outcomes for survival and competing risks
analysis based on \color{black} a specialization of the work by
Rytgaard et al.  (2020), and, further, derive exact expressions for
the second-order remainders to establish double robustness properties
of the considered survival and competing risks estimation problems.

Generally, targeted maximum likelihood estimation proceeds in two
steps.  Initial estimators are constructed for the nuisance parameters
in the first step which is then followed by a second step (the
targeting step) that reduces bias for the initial estimators, improves
precision and ensures reliable statistical inference in terms of
confidence intervals and p-values.  We suggest a targeting step based
on a proportional hazards type submodel for the intensity of the event
process, iteratively updating the hazard estimators carried out in a
smoothed fashion across time until the optimal score equation is
solved.  We further construct a flexible highly adaptive lasso
estimator \citep{van2017generally} for continuous-time hazards that
can be implemented with \(L_1\)-penalized regression
\citep{tibshirani1996regression} by utilizing Poisson regression
modeling techniques
\citep{andersen2012statistical,lindsey1995fitting}.  The highly
adaptive lasso is a nonparametric estimation method proven to converge
faster than \(n^{-1/4}\) with respect to the square-root of the
loss-based dissimilarity under the only assumption that the true
function is \cadlag (right-continuous with left limits) with a finite
sectional variation norm \citep{van2017generally}.

Our estimation methods have applications in observational as well as
experimental data settings such as randomized clinical trials. In
trial settings, it is standard to analyze time-to-event data in each
intervention arm using a Kaplan-Meier estimator. It is well-known that
the Kaplan-Meier estimator, an unadjusted estimator, may yield biased
estimation under covariate dependent censoring, and may, furthermore,
be inefficient in presence of predictive covariates
\citep{lu2008improving,rubin2008empirical,moore2009increasing,rotnitzky2012improved,diaz2019improved}. We
revisit these issues in our simulation study, where we demonstrate
that our targeted maximum likelihood estimator based on the highly
adaptive lasso estimator improves the precision and robustness of
findings while providing accurate confidence intervals without any
parametric model specification.

\section{Setting and notation}
\label{sec:actual:setting:notation}

Consider unit-specific observed data on the form
\(O= (L, A, \tilde{T}, \Delta)\). These data contain a vector of
information on pre-treatment characteristics
(\(L\in \mathcal{L}=\R^d\)), tell us what treatment the unit was
exposed to at study entry (\(A\in \mathcal{A}= \lbrace 0,1\rbrace\)),
how long the unit was under observation (\(\tilde{T}\in \R_+\)) and if
the event of interest was observed for the unit or not
\((\Delta \in \lbrace 0,1\rbrace\)).  We introduce the variables
\(T\in \R_+\) and \(C\in \R_+\) as the times to event and censoring,
respectively, such that the observed time is
\(\tilde{T} = \min ( T, C)\) and the event indicator can be written
\(\Delta = \1\lbrace T \le C\rbrace \).  The corresponding competing
risks setting is treated in Appendix A.  Suppose that we observe a
sample \(\lbrace O_i\rbrace_{i=1}^n\) of independent and identically
distributed observations of \(O\sim P_0 \). Let
\(N(t) = \1\lbrace \tilde{T} \le t, \Delta=1\rbrace\) and
\(N^c(t) = \1\lbrace \tilde{T} \le t, \Delta=0\rbrace\) denote the
observed counting processes \citep{andersen2012statistical}.  \iffalse
A useful representation of the observed data is given in terms of
counting processes \citep{andersen2012statistical}. Specifically,
define the counting processes
\(N(t) = \1\lbrace \tilde{T} \le t, \Delta=1\rbrace\) and
\(N^c(t) = \1\lbrace \tilde{T} \le t, \Delta=0\rbrace\) and denote by
\(\Lambda_0^d, \Lambda_0^c \) the cumulative intensities
characterizing the compensator of \(N, N^c\), respectively. Assuming
that \(\Lambda_0, \Lambda_0^c \) are absolute continuous, \fi The
hazards for the distributions of \(T\) and \(C\) are defined by
\begin{align*}
  \lambda_0(t \, \vert \, a, \ell)
  & =
    \underset{h \rightarrow 0}{\lim} \,\, h^{-1} P(T \le t+h \mid \tilde{T} \ge t, A=a, L=\ell), \\
  \lambda^c_0(t \, \vert \, a, \ell)
  & =
    \underset{h \rightarrow 0}{\lim} \,\, h^{-1} P(C \le t+h \mid \tilde{T} \ge t, A=a, L=\ell) ; 
\end{align*}
let \(\Lambda_0, \Lambda_0^c \) denote the corresponding cumulative
hazards; thus, the compensators of \(N, N^c\) are characterized by
\begin{align*}
  \EE_{P_0} [ N(dt) \mid \F_{t-}] = \1 \lbrace \tilde{T}\ge t\rbrace \Lambda_0(dt \, \vert \, A, L),
  \,\, \text{and,} \,\,
  \EE_{P_0} [ N^c(dt) \mid \F_{t-}] = \1 \lbrace \tilde{T}\ge t\rbrace \Lambda^c_0(dt \, \vert \, A, L),
\end{align*}
where \(\F_{t} = \sigma ((N(s) , N^c(s) \, : \, s \le t), A , L)\) is
the filtration generated by the observed data in \([0,t]\).  Let
further \(\mu_0\) be the density of \(L\) with respect to an
appropriate dominating measure \(\nu\) and
\(\pi_0(\cdot \, \vert \, L) \) be the conditional distribution of
\(A\) given \(L\). Under coarsening at random
\citep{vanRobins2003unified}, \(\lambda, \lambda^c\), the distribution
for the observed data can now be represented as
\( dP_0(o) = p_0(o) d\nu(\ell) dt\) where the density \(p_0\) is given
by:
\begin{align}
  \begin{split}
    p_0(o) = \mu_0(\ell) \pi_0(a\, \vert \, \ell) \big(
    \lambda_0 (t \mid a, \ell) \big)^{\delta}S_0(t - \mid a,
    \ell) \big( \lambda^c_0 (t \mid a, \ell)
    \big)^{1-\delta} S_0^c(t - \mid a, \ell);
\end{split}
  \label{eq:survival:factorization}
\end{align}
here \(o = (\ell, a, t, \delta)\) and \(S_0\), \(S^c_0\) denote
\begin{align}
  S_0 (t \mid a, \ell)
  & =\Prodi_{s \le t} \big( 1- \Lambda_0 (ds \mid a, \ell) \big) = 
    \exp \Big( - \int_0^t \lambda_0(s \mid a, \ell)ds\Big), \label{eq:def:survival} \\ 
  S^c_0 (t \mid a, \ell)
  &=\Prodi_{s \le t} \big( 1- \Lambda^c_0 (ds \mid a, \ell) \big) =
    \exp \Big( - \int_0^t \lambda^c_0(s \mid a, \ell)ds\Big), \label{eq:def:cens:survival}
\end{align}
the survival functions for \(T\) and \(C\), respectively. In the above
display, \( \prodi\) denotes the product integral
\citep{gill1990survey,andersen2012statistical}.

\section{Target of estimation}
\label{sec:interventions:post:distr}

We formulate our target parameter under hypothetical interventions on
the baseline treatment decision governed by \(\pi_0\) and the
censoring mechanism governed by \(\lambda_0^c\). For this
purpose, we define:
\begin{align}
  g_{0} (o) & =   
              \pi_0(a\, \vert \, \ell) \big( \lambda^c_0 (t \mid a, \ell) \big)^{1-\delta}
              S_0^c(t -\mid a, \ell), 
              \label{eq:G:part} \\
  q_{0} (o)& =  \mu_0(\ell)
             \big( \lambda_0 (t \mid a, \ell)\big)^{\delta}
             S_0(t -\mid a, \ell). 
  \label{eq:Q:part}
\end{align}
We refer to \eqref{eq:G:part} as the \textit{interventional} and to
\eqref{eq:Q:part} as the \textit{non-interventional} part of the
likelihood. The distribution \(P_0 \) that factorizes as in
\eqref{eq:survival:factorization} can now be parametrized by
\eqref{eq:G:part}--\eqref{eq:Q:part}: We write \(P_{q_0,g_0}\). We
consider the statistical model \( \mathcal{M}\) for \(P_0\) consisting
of distributions that admit an equivalent parametrization:
\begin{align*}
  \mathcal{M} = \Big\lbrace P \, : \, P= P_{q,g} , \, q
  \in \mathcal{Q}, g\in \mathcal{G}\Big\rbrace,
\end{align*}
with parameter space \(\mathcal{Q}\) for the non-interventional part
and \(\mathcal{G}\) for the interventional part. An intervention
\(g \mapsto g^*\) is imposed on \(P\in\mathcal{M}\) by substituting
\(g^*\) for \(g\), so that \(P_{q,g^*}\) defines the
post-interventional distribution, also known as the g-computation
formula \citep{robins1986new}.  Such an intervention \(g^*\) can be
specified as follows:
\begin{align}
  g^*(o) = \delta \,\pi^*(a \mid \ell)
  ,
  \label{eq:G:a:0}
\end{align}
imposing the treatment decision \(\pi^*\) at baseline and no censoring
throughout the follow-up period. To define the average treatment
effect, for example, we specify the interventions \(g^{a^*}\),
\(a^*=0,1\), where
\begin{align}
  g^{a^*}(o) = \delta \,\1 \lbrace a=a^*\rbrace,
  \label{eq:G:a}
\end{align}
i.e., \(\pi^* \) is a degenerate distribution that puts all mass in
\(A=a^*\), \(a^*=0,1\).

With an intervention \(g^*\) and corresponding post-interventional
distribution \(P_{q,g^*}\) as defined in Section
\ref{sec:interventions:post:distr}, we now define our target parameter
\(\Psi_{\tmax} \, : \, \mathcal{M} \rightarrow \R\) as the
intervention-specific mean outcome:
\begin{align}
  \Psi_{\tmax}(P) = \EE_{P_{q,g^*}}  [\1 \lbrace \tilde{T} \le \tmax\rbrace  ] =
  1-\int_{\mathcal{L}} \sum_{a=0,1}
  S(\tmax \mid a, \ell) \pi^*(a \mid \ell) \mu( \ell) d\nu (\ell)
  ,
  \label{eq:target:parameter}
\end{align}
under imposing the hypothetical intervention \(g^*\) on the
distribution \(P\in \mathcal{M}\). We denote by
\(\psi_0 = \Psi_{\tmax}(P_0)\) the true value. Under causal
assumptions reviewed in Remark \ref{remark:target:parameter:causal}
below, the target parameter can be interpreted as the risk we would
observe had subjects been treated according to the treatment strategy
\(\pi^*\).  Importantly, the average treatment effect can be defined
in terms of a contrast between two intervention-specific mean
outcomes, namely:
\begin{align}
  \Psi^{\mathrm{ATE}}_{\tmax}(P) =  \Psi^{1}_{\tmax}(P) - \Psi^{0}_{\tmax}(P) ,
  \quad \text{ where, } \quad  \Psi^a_{\tmax}(P) =  \EE_{P_{q,g^a}} [\1 \lbrace \tilde{T}\le \tmax\rbrace  ], \text{ for }
  a=0,1,
  \label{eq:target:parameter:ate}
\end{align}
which, under the causal assumptions listed below, is interpreted as
the average treatment effect on the \(\tmax\)-year risk had we
randomized subjects to treatment or no treatment.

\begin{remark}[Causal interpretation]
  \label{remark:target:parameter:causal}
  Define the hypothetical event time \(T^{g^*} \) had the subject been
  treated according to the treatment strategy \(\pi^*\), irrespective
  of what it actually was.  Particurly, let \(T^a\) be the
  hypothetical event time that would result if the treatment decision
  had been \(A = a\).  We formulate the causal assumptions as follows.

  \begin{assumption}[Consistency]
    \(T=T^a\) on the event that \(A=a\), for \(a=0,1\).
    \label{ass:A1}
    \end{assumption}

  \begin{assumption}[Conditionally independent censoring]
    \(T\independent C \,\vert \, A, L\).
    \label{ass:A2}
  \end{assumption}

  \begin{assumption}[No unmeasured confounding]
    \(T^a \independent A \,\vert \, L \), for \(a=0,1\).
    \label{ass:A3}
  \end{assumption}

  \begin{assumption}[Positivity]
    \(P(C \ge \tmax \mid a, L) \pi(a \mid L) > \eta > 0 \), a.e., for
    \(a=0,1\).
    \label{ass:A4}
  \end{assumption}
  
  Under Assumptions \ref{ass:A1}--\ref{ass:A4}, the target parameter
  can be interpreted as the \(\tmax\)-year risk we would observe had we in
  fact imposed the intervention \(\pi^*\) in the real world, i.e.,
  \begin{align}
    \Psi_{\tmax} (P)  = P ( T^{g^*}  \le \tmax).
    \label{eq:causal:interpreation}
  \end{align}
  Particularly, the average treatment effect, defined by Equation
  \eqref{eq:target:parameter:ate}, is equal to
  \( \Psi^{\mathrm{ATE}}_{\tmax}(P) = P( T^1 \le \tmax ) - P( T^0 \le \tmax
  )\).\\
  \label{remark:causal}
\end{remark}
We show in the supplementary material that
\eqref{eq:causal:interpreation} follows under Assumptions
\ref{ass:A1}--\ref{ass:A3}.

\section{Efficient estimation}
\label{sec:efficient:estimation}

An estimator \( \hat{\psi}^*_n = \Psi_{\tmax} (\hat{P}_n^*)\) for the
target parameter \(\psi_0\) is obtained by providing an estimator
\(\hat{P}_n^*\) for (the relevant components of) \(P_0\) and plugging
this into the parameter mapping \eqref{eq:target:parameter}. The
target parameter only depends on \(P\) through \(\lambda\) and
the marginal distribution \(\mu\) of \(L\). We will estimate the
latter by the empirical mean so that an estimator for \(\psi_0\) can
be based solely on an estimator \(\hat{\lambda}\) for
\(\lambda\). Particularly, an estimator
\(\hat{{\lambda}}_{n}\) is mapped to an estimator \(\hat{{S}}_{n}\)
for the conditional survival function by evaluating Equation
\eqref{eq:def:survival}, and next to an estimator for the target
parameter by:
\begin{align}
  1- \frac{1}{n} \sum_{i=1}^n \bigg(
  \sum_{a=0,1} \hat{S}_{n}( \tmax \mid a, L_i)  \pi^*(a \mid L_i) \bigg)
  .
  \label{eq:construct:est:0}
\end{align}
Efficient estimation, on the other hand, also requires estimation of
the interventional part of the likelihood. Here, and throughout,
\(\mathbb{P}_n\) denotes the empirical distribution of the data
\(\{O_i\}_{i=1}^n\) and \(o_P(1)\) is a term that converges to zero in
probability. An estimator \(\hat{\psi}_n \) for the target parameter
is asymptotically linear with influence curve equal to the efficient
influence curve \(D_{\tmax}^*(P_0)\)
\citep{bickel1993efficient,van2000asymptotic} if and only if
\begin{align}
  \sqrt{n} \big(  \hat{\psi}_n - \psi_0 \big) = \sqrt{n} \, \mathbb{P}_n D^*_{\tmax}(P_0) + o_P(1). 
\label{eq:efficient:estimator}
\end{align}
The efficient influence function \(D_{\tmax}^*(P)\) for our statistical
model \(\mathcal{M}\) and target parameter
\(\Psi_{\tmax} \, : \, \mathcal{M} \rightarrow \R\) is well-known from
the literature
\citep{robins1992recovery,vanRobins2003unified,moore2009application,moore2009covariate,van2011targeted}. A
sketch of its derivation can be found in the supplementary material;
we present an expression for \(D_{\tmax}^*(P)\) in Section
\ref{sec:tmle}. Specifically, the efficient influence function is a
mapping of \(P\in\mathcal{M}\) through \(\pi\), \(\lambda^c\)
and \(\lambda\), thus comprising the nuisance parameters for
our estimation problem.

\subsection{Conditions for asymptotically linear and efficient
  estimation}
\label{sec:conditions:for:al:eff}

Below we give sufficient conditions for establishing
\eqref{eq:efficient:estimator}; the proof follows similar work
\citep[see, e.g.,][Theorem
A.5]{van2006targeted,van2017generally,van2011targeted} but is included
for completeness in the supplementary material. Define the
second-order remainder:
\begin{align}
  R(P,P_0):= \Psi_{\tmax} ( P ) - \Psi_{\tmax} (P_0 ) + P_0 D_{\tmax}^*(P), \qquad P \in \mathcal{M}.
  \label{eq:def:R2}
\end{align}
Consider an estimator \(\hat{P}^*_n\) for \(P_0\) which solves the
efficient influence curve equation: 
\begin{align}
  \mathbb{P}_n D^*_{\tmax}(\hat{P}^*_n
  ) = o_P(n^{-1/2}).
  \label{eq:key:EIC:eq}
\end{align}
Now, if the following conditions (i) and (ii) hold true,
\begin{enumerate}
\item[(i)] \(R(\hat{P}^*_n, P_0 ) = o_P(n^{-1/2})\),\vspace{0.2cm}
\item[(ii)] \(D^*_{\tmax}(\hat{P}^*_n )\) belongs to a Donsker class,
  and
  \(P_0\big( D^*_{\tmax}(\hat{P}^*_n ) - D^*_{\tmax}(P_0 )\big)^2\)
  converges to zero in probability,
\end{enumerate}
then \eqref{eq:efficient:estimator} holds for
\(\hat{\psi}^*_n = \Psi_{\tmax} (\hat{P}^*_n)\), that is,
\( \Psi_{\tmax} (\hat{P }^*_n)\) is asymptotically linear at \(P_0\)
with influence curve \(D^*_{\tmax}(P_0)\).

To construct an efficient estimator for the target parameter, the
nuisance parameters \(\pi\), \(\lambda^c\) and \(\lambda\) must be
estimated such that the conditions above are met.  To shed light on
conditions (i) and (ii), we derive the exact second-order remainder
\(R(P,P_0)\) as displayed in Section \ref{remark:R2} below; in the
subsequent Section \ref{eq:donsker:etc} we present the overall
smoothness restrictions on the statistical model (Assumption
\ref{ass:overall:cadlag}) that combined with highly adaptive lasso
estimation provides the basis for establishing conditions (i) and
(ii). \color{black}

\subsection{Double robustness of the second-order remainder}
\label{remark:R2}

As we show in the supplementary material, the second-order remainder
defined in Display \eqref{eq:def:R2} can be written out explicitly as
follows:
 \begin{align*}
   R(P,P_0)  
   =  \int_{\mathcal{O}} \1 \lbrace t \le \tmax\rbrace  \left( \frac{
   {S}_0^c(t - \mid a,\ell) \pi_0 (a \mid \ell)  -  {S}^c(t- \mid a,\ell)
   \pi  (a\mid \ell)}{ S^c(t- \mid a,\ell) \pi (a \mid \ell)}\right)   \\
   {S}_0 (t- \mid a,\ell)
   \big(
   {\lambda}_0( t \mid a,\ell)  -
   {\lambda}( t \mid a,\ell)  \big) dt \\
   \frac{{S}(\tmax \mid a, \ell)}{{ S}(t \mid a,\ell)}
   \pi^*(a \mid \ell) \color{black}\,
   \mu_0 (\ell) d\nu (\ell).
\end{align*}
Importantly, this second-order remainder displays a double robustness
structure: We see that \(R(\hat{P}_n,P_0)=0\) if either
\begin{enumerate}
\item \(\pi_0 (a \mid \ell)= \hat{\pi}_n (a \mid \ell)\) and
  \({S}_0^c( t \mid a, \ell) = \hat{S}^c_n ( t \mid a, \ell)\) for all \(t \in (0, \tmax)\), \\[-0.5em]
\item[]  or,  \\[-0.5em]
\item \({\lambda}_0( t \mid a,\ell) = \hat{\lambda}_n( t \mid a,\ell) \) for
  all \(t \in (0,\tmax)\).
\end{enumerate}
Let \(G (t \mid a, \ell):= {S}^c(t- \mid a,\ell)\pi(a\mid \ell)\).
When \(G \) is bounded away from zero by some \(\eta >0\) (Assumption
\ref{ass:A4}), the denominator of the first factor of \( R(P,P_0) \)
is bounded from above by \(\eta^{-1} \). Furthermore,
\(S_0(t- \mid a,\ell)\) is bounded by 1, and, since \(t\le \tmax\), we
have that \(S(\tmax \mid a,\ell) / S(t \mid a,\ell) \le 1\).  Now, the
product structure of the remainder \( R(P,P_0) \) yields, by the
Cauchy-Schwartz inequality an upper bound as follows:
\begin{align*}
  & \vert  R(P,P_0) \vert \\
  &\quad \le 
    \eta^{-1}   \int_{\mathcal{L}} \sum_{a=0,1} \int_0^{ \tmax} \big(
    G_{0} (t\mid a, \ell)  -  G (t \mid a, \ell)
    \big)   
    \big(
    {\lambda}_{0} (t\mid a, \ell)  - 
    {\lambda} (t \mid a, \ell )\big) dt  \pi^* (a \mid \ell)d\mu_0 (\ell) \notag \\
  &\quad \le \eta^{-1}    \big\Vert
    G_{0}  -  G_{}
    \big\Vert_{\pi^* \otimes \mu_0\otimes \rho}
    \big\Vert
    {\lambda}_{0}   - 
    {\lambda}   \big\Vert_{\pi^* \otimes \mu_0\otimes \rho} ,
  \end{align*}
  where
  \(\Vert f\Vert_{\pi^* \otimes \mu_0\otimes \rho} = \sqrt{ \int f^2
    d(\pi^* \otimes \mu_0\otimes \rho)}\) and \(\rho\) denotes the
  Lebesgue measure on \([0,\tmax]\). We see that the required
  convergence rate, \(R (\hat{P }^*_n , P _0 ) = o_P(n^{-1/2})\), for
  example is achieved if we estimate both \(G_{0}\) and
  \(\lambda_{0}\) at a rate faster than \(o_P(n^{-1/4})\) with respect
  to the \(L_2(\pi_0 \otimes \mu_0\otimes \rho)\)-norm.

\subsection{Efficient estimation under weak conditions on
  \(\mathcal{M}\)}
\label{eq:donsker:etc}

As we sketch below, the following weak conditions assumed for the
statistical model \(\mathcal{M}\) combined with the use of highly
adaptive lasso estimation (Section \ref{sec:initial:estimation}) for
the nuisance parameters provides the basis for conditions (i) and (ii)
of Section \ref{sec:conditions:for:al:eff}.

\begin{assumption}[Conditions on \(\mathcal{M}\)]
  Assume that the nuisance parameters \(\lambda,\lambda^c, \pi\) can
  be parametrized by functions that are \cadlag and have finite
  sectional variation norm
  \citep{gill1995inefficient,van2017generally}, that positivity holds
  (Assumption \ref{ass:A4}) and further that \(S(\tmax)>\eta'\) for
  some \(\eta'>0\).
  \label{ass:overall:cadlag}
\end{assumption}

Assumption \ref{ass:overall:cadlag} has the following two important
implications. First, the class of \cadlag functions with finite
variation is a Donsker class \citep{van1996weak}. Since the efficient
influence function is a well-behaved mapping of the nuisance
parameters, it inherits the Donsker properties. Second, it constitutes
the basis for construction of the highly adaptive lasso estimator
shown to converge at a rate faster than \(n^{-1/4}\) to its true
counterpart belonging to the class of \cadlag functions with finite
sectional variation \citep{van2017generally}.  The convergence is with
respect to the square-root of the loss-based dissimilarity which
behaves as the \(L_2(\pi_0 \otimes \mu_0\otimes \rho)\)-norm (Appendix
C).  In light of the double robustness properties of the second-order
remainder (Section \ref{remark:R2}), we see that Assumption
\ref{ass:overall:cadlag} combined with highly adaptive lasso
estimation provides the basis for establishing conditions (i) and (ii)
of Section \ref{sec:conditions:for:al:eff}.

\section{Targeting algorithm}
\label{sec:tmle}

Suppose we have at hand estimators \(\hat{\pi}_n\),
\(\hat{\lambda}^c_n\) and \(\hat{\lambda}_{n}\). The overall idea of
targeted maximum likelihood estimation is to perform an update of
\(\hat{\lambda}_n \mapsto \hat{\lambda}_{n,*}\), for the given
estimators \(\hat{\pi}_n\), \(\hat{\lambda}^c_n\), such as to solve
the efficient influence curve equation. The estimator
\(\hat{\lambda}_{n,*}\) is then mapped to an estimator
\(\hat{S}_{n,*}\) according to \eqref{eq:construct:est:0} to
construct an estimator for the target parameter.

In the following Definitions \ref{defi:clever:weigths} and
\ref{defi:clever:covariates}, we define so-called `clever weights' and
`clever covariates' that will allow for a concise representation of
the efficient influence function that we utilize in the construction
of our targeting algorithm.

\begin{definition}[Clever weights]
\label{defi:clever:weigths}
Define clever weights by:
\begin{align*}
  w_t(O) & = \1 \lbrace \tilde{T} \ge t \rbrace 
           \frac{
           \pi^*(A \mid L) }{ \pi (A \, \vert \, L) } \frac{\1 \lbrace t  \le \tmax \rbrace}{
           {S}^c( t- \, \vert \,A, L)
           }, \quad t >0. 
\end{align*}
\end{definition}

Notably, the clever weights depend on the interventional part of the
data-generating distribution and on the choice of intervention
\(\pi^*\). In our simulation study (Section
\ref{sec:simulation:study}), we focus on the average treatment effect
displayed in \eqref{eq:target:parameter} which we target directly. In
this case the clever weights are defined in terms of both
interventions \(\pi^1(A \mid L) = \1\lbrace A=1\rbrace\) and
\(\pi^0(A \mid L) = \1\lbrace A=0\rbrace\) as follows:
\begin{align}
  w^{\mathrm{ATE}}_t(O) & =  
                          \1 \lbrace \tilde{T} \ge t \rbrace   \left(  \frac{
                          \1 \lbrace A=1 \rbrace}{ \pi (1 \, \vert \, L) }
                          -
                          \frac{
                          \1 \lbrace A=0 \rbrace}{ \pi (0 \, \vert \, L) } \right)
                          \frac{\1 \lbrace t  \le \tmax \rbrace}{
                          {S}^c( t- \, \vert \, A, L)
                          }.
                          \label{eq:clever:weights:ate}
\end{align} 
The clever covariates defined below, to the contrary, only depend on
the non-interventional part and are thus fixed across choices of
interventions. Estimators for the clever weights remain constant,
whereas estimators for the clever weights are updated as part of the
targeting algorithm.

\begin{definition}[Clever covariates]
  \label{defi:clever:covariates}
  Define clever covariates by:
  \begin{align*}
    h_t (O) & = 
              \frac{ {S}(\tmax \, \vert \, A,L)}{
             {S}(t \, \vert \, A,L)
              }, \quad t>0 
              .
  \end{align*}
\end{definition}

With \(w_t\) and \(h_t\) defined by Definitions
\ref{defi:clever:weigths} and \ref{defi:clever:covariates}, the
efficient influence function for our statistical model \(\mathcal{M}\)
and target parameter \(\Psi_{\tmax} \, : \, \mathcal{M} \rightarrow \R\)
can now be represented on the concise form:
\begin{align}
  \begin{split}
    D_{\tmax}^*(P) (O) &= \int_0^{\tmax} {w}_t (O) h_t (O) \, \big(
    N(dt) -
    {\Lambda}(dt \, \vert \, A,L)\big)  \\
    & \qquad\qquad\qquad\qquad + \, 1- \sum_{a=0,1} {S}(\tmax \, \vert
    \, a, L) \pi^*(a \mid L) - \Psi_{\tmax}(P) .
    \end{split}
\label{eq:surv:eff:ic:continuous:1}
\end{align}

Any estimator on the form \eqref{eq:construct:est:0} solves all but
the first term of the efficient influence curve equation, so what
remains is the first term, corresponding to the first term of
\eqref{eq:surv:eff:ic:continuous:1}, giving rise to the equation:
\color{black}
\begin{align}
  \mathbb{P}_n D^*_{1,\tmax} (\hat{P}_n) = o_P(1), \,\quad \text{where,} \quad
  D^*_{1,\tmax}(P) (O):=  \int_0^{\tmax} {w}_t (O)   h_t (O) \, \big( N(dt) -
  {\Lambda}(dt \, \vert \, A,L)\big) .
  \label{eq:eff:ice:1}
\end{align}

Our algorithm is formed by iterative update steps for
\(\hat{\lambda}_n\) along a path defined by a one-dimensional
fluctuation model, defining a sequence of estimators,
\begin{align}
  \hat{\lambda}_n = \hat{\lambda}_{n,k=0}, \quad
  \hat{\lambda}_{n,k=1}, \quad \hat{\lambda}_{n,k=2}, \quad
  \ldots 
  \label{eq:seq:estimators}
\end{align}
such that \(\hat{P}_n^*\), characterized by \(\hat{\lambda}^c_n\),
\(\hat{\pi}_n\) and a final estimator
\( \hat{\lambda}_{n,*}= \hat{\lambda}_{n,k=k^*}\) from the sequence
\eqref{eq:seq:estimators}, solves \eqref{eq:eff:ice:1}. We refer to
\( \hat{\lambda}_n = \hat{\lambda}_{n,k=0}\) as the \textit{initial
  estimator} and to \( \hat{\lambda}_{n,*}= \hat{\lambda}_{n,k=k^*}\)
as the \textit{targeted estimator}.

\subsection{Fluctuation model and update algorithm for \({\lambda}\)}
\label{ssec:loss:submodel:intensities}

The update steps for each element \( \hat{\lambda}_{n,k}\) of the
sequence of estimators \eqref{eq:seq:estimators} are performed along a
one-dimensional fluctuation model through \(\hat{\lambda}_{n,k}\). To
this end, we define the multiplicative hazards type fluctuation model:
\begin{align}
  \begin{split}
    \lambda (t; \eps) & = \lambda(t) \exp ( \eps {w}_t {h}_{t}), \qquad \eps \in \R. 
\end{split}
  \label{eq:lambda:fluc}
\end{align}
The estimators \(\hat{\pi}_n\) and \(\hat{\lambda}^c_n\) together
define an estimator \( \hat{w}_{t}\) for the clever weights, whereas
\(\hat{\lambda}_{n}\) defines an estimator \(\hat{h}_{t}\) for the
clever covariate.  Now, plugging the evaluation of
\eqref{eq:lambda:fluc} in the current estimator
\(\hat{\lambda}_{n,k}\), the estimator \( \hat{w}_{t}\) for the
clever weights and the estimator \(\hat{h}_{t,k}\), obtained from
\(\hat{\lambda}_{n,k}\), for the clever covariate into
\eqref{eq:eff:ice:1} defines an equation in \(\eps\):
\begin{align}
  \begin{split}
    &0= \frac{1}{n} \sum_{i=1}^n \Bigg(
    \int_0^{\tmax}  \hat{w}_t (O_i) \hat{h}_{t,k} (O_i)\, N_i(dt)  \\
    &\qquad\qquad - \int_0^{\tmax} \hat{w}_t (O_i)\hat{h}_{t,k} (O_i)
    \exp (\eps \hat{w}_t (O_i) \hat{h}_{t,k} (O_i) ) \,
    \hat{\lambda}_{n,k} (t \mid A_i, L_i ) dt \Bigg) .
    \end{split}\label{eq:equation:written:out:0}
\end{align}
The solution \(\hat{\eps}_k\) to \eqref{eq:equation:written:out:0}
defines the update of \(\hat{\lambda}_{n,k}\) along the fluctuation
model: 
\begin{align*}
  \hat{\lambda}_{n,k+1} := \hat{\lambda}_{n,k}(t) \exp ( \hat{\eps}_k \hat{w}_t \hat{h}_{t,k}),
  \quad
  k \ge 0 . 
\end{align*}
The steps from \(k\) to \(k+1\) are repeated until
\(\hat{\eps}_{k^*}\approx 0\), or, more precisely, the corresponding
estimator \(\hat{P}_n^*\), consisting of \(\hat{\lambda}^c_n\),
\(\hat{\pi}_n\) and \( \hat{\lambda}_{n,*} = \hat{\lambda}_{n,k^*}\),
solves
\begin{align*}
  \vert \, \mathbb{P}_n D_{\tmax}^*(\hat{P}_n^{*}) \, \vert \le s_n,
\end{align*}
for the stopping criterion
\(s_n = \hat{\sigma}_n / (n^{-1/2} \log n) \), where
\(\hat{\sigma}_n^2\) is the estimated variance of the efficient
influence function.

\subsection{Final estimation}
\label{ssec:tmle:final}

The final estimator \(\hat{\lambda}_{n,*}\) defines a corresponding
estimator \(\hat{S}_{n, *}\) for the survival function based on which
we can construct an estimator of the target parameter by:
\begin{align*}
  \hat{\psi}^*_n = 1-\frac{1}{n}\sum_{i=1}^n
  \bigg(
  \sum_{a=0,1} \hat{S}_{n,*}( \tmax \mid a, L_i)  \pi^*(a \mid L_i) \bigg)
  . 
\end{align*}
Under conditions (i) and (ii) of Section \ref{sec:conditions:for:al:eff}, we
can use the asymptotic normal distribution
\begin{align*}
  \sqrt{n}\, \big( \hat{\psi}^*_n
  - \psi_0 \big)
  \overset{\mathcal{D}}{\rightarrow} \mathcal{N} (0, {P}_0  D^*_{\tmax} ( {P}_0
  )^2), 
\end{align*}
to provide an approximate two-sided confidence interval. The
asymptotic variance of the estimator is equal to the variance of the
efficient influence function and can be estimated by
\(\hat{\sigma}_n^2 = \mathbb{P}_n (D^*_{\tmax} ( \hat{P}^*_n))^2\).

\section{Initial estimation}
\label{sec:initial:estimation}

To carry out our targeting algorithm, we need initial estimators for
the conditional hazards \(\lambda^c\) and \(\lambda\) of the censoring
process and the event process, respectively, and further for the
conditional distribution of treatment given covariates
\(\pi\). Estimation of \(\pi\) can be done by any binary regression
method, including logistic regression and a large variety of machine
learning algorithms. Thus, our focus is here on the estimation of a
general continuous-time conditional hazard, denoted
\( \lambda (t \, \vert \, Z) \), where, for our purposes, \(Z\)
consists of the baseline treatment and covariates,
\(Z=(A, L)\in \R^{d+1}\). Particularly, following previous work
\citep[][Chapter
6,7]{benkeser2016highly,van2017generally,van2018targeted}, Section
\ref{sec:poisson:hal} below presents our highly adaptive lasso
estimator for such continuous-time conditional hazards.

\subsection{Highly adaptive lasso estimation of hazards}
\label{sec:poisson:hal}

To construct our highly adaptive lasso estimator for the conditional
hazard \( \lambda (t \, \vert \, Z) \), we propose the
reparametrization as follows:
\begin{align}
  \lambda (t \, \vert \, Z)  =  
  \exp ( f(t, Z) ), \qquad f \, : \, [0,\tmax] 
  \times \R^{d+1} \rightarrow \R, 
  \label{eq:parametrization}
\end{align} 
and denote by \([0,\kappa]\subset\R^{d+1}\) the support of
\( z \mapsto f(t, z)\) with \(\kappa = \infty\) allowed. The steps to
define the highly adaptive lasso estimator for \(\lambda\) involves
some technical parts following earlier work \citep{van2017generally},
and, specifically, we need notation for the sectional variation of
\(f\) that we now present. For a subset of indices
\(\mathcal{S}\subset \lbrace 1,\ldots, d+1\rbrace\), we denote by
\(z_{\mathcal{S}} \) the \(\mathcal{S}\)-specific coordinates of
\(z\in \R^{d+1}\) and by
\( z \mapsto f_{\mathcal{S}} (t,z) = f(t, z_{\mathcal{S}},
0_{\mathcal{S}^c})\) the \(\mathcal{S}\)-specific section of \(f\)
that sets the coordinates in the complement of \(\mathcal{S}\) equal
to zero. As we state below, we will make the assumption that the
function \(f(t,z)\) is \cadlag (i.e., right-continuous with left
limits), then, following \cite{gill1995inefficient,van2017generally},
the sectional variation norm of \(f\) is
\begin{align*}
  \Vert f \Vert_{\vv}
  &= \vert f(0,0) \vert + \int_{(0, \tmax]} \vert    f (dt, 0) \vert +
    \sum_{\mathcal{S}\subset \lbrace 1, \ldots, d+1\rbrace}
    \bigg( 
    \int_{(0_{\mathcal{S}}, \kappa_{\mathcal{S}}]} \vert  f_{\mathcal{S}} (0, dz) \vert \\
  &\qquad\qquad\qquad\qquad
    \qquad\qquad\qquad\qquad\qquad\qquad + 
    \int_{(0, \tmax]}\int_{(0_{\mathcal{S}}, \kappa_{\mathcal{S}}]} \vert f_{\mathcal{S}} (dt, dz) \vert
    \bigg) 
    .
\end{align*}
The key to defining the highly adaptive lasso estimator is the
following assumption on the function class containing \(f\).  Let
\(\mathscr{F}_{\!\! \mathscr{M}}\) denote the class of \cadlag
functions with sectional variation norm bounded by a constant
\(\mathscr{M}<\infty\). We assume that
\(f\in \mathscr{F}_{\!\! \mathscr{M}} \).

Highly adaptive lasso estimation of \(f\) is defined by the
infinite-dimensional minimization problem over all
\(f \in \mathscr{F}_{\!\! \mathscr{M}}\):
\begin{align}
  \underset{f \in \mathscr{F}_{\!\! \mathscr{M}} }{\min} \mathbb{P}_n 
  \mathscr{L}(f) ,
  \label{eq:definition:hal:ideal}
\end{align}
for a loss function \((O,f) \mapsto \mathscr{L}(f) (O)\). The
practical construction consists of approximating the minimizer over
all \(\mathscr{F}_{\!\! \mathscr{M}}\) in
\eqref{eq:definition:hal:ideal} by the minimizer over discrete
measures in \(\mathscr{F}_{\!\!\mathscr{M}}\).  Particularly, the
assumption that \(f\in \mathscr{F}_{\!\! \mathscr{M}} \) yields a
representation for \(f\in \mathscr{F}_{\!\!\mathscr{M}}\) in terms of
its measures over sections \citep{gill1995inefficient}
\begin{align*}
  f (t,z)
  &=  f(0,0)  + 
    \int_{(0, \tmax]} \1\lbrace s \le t\rbrace  f (ds,0) +   \sum_{\mathcal{S}\subset \lbrace 1, \ldots, d+1\rbrace}
    \bigg(
    \int_{(0_{\mathcal{S}}, \kappa_{\mathcal{S}}]}  
    \1 \lbrace u \le z_{\mathcal{S}}\rbrace  f_{\mathcal{S}} ( 0,du)  \\
  &\qquad\qquad\qquad\qquad
    \qquad\qquad\qquad + 
    \int_{(0, \tmax]}\int_{(0_{\mathcal{S}}, \kappa_{\mathcal{S}}]} \1\lbrace s \le t\rbrace
    \1 \lbrace u \le z_{\mathcal{S}}\rbrace
    f_{\mathcal{S}} (ds, du) 
    \bigg) 
    , 
\end{align*} 
which for an approximation over a finite support becomes a finite
linear combinations of indicators functions and corresponding
coefficients being the pointmass assigned to support points.

Let us consider a grid of time-points partitioning \([0, \tmax]\),
\( 0=t_0 <t_1 <\cdots <t_R <t_{R+1}= \tmax \), and further a
partitioning \(\mathcal{Z}_1 \cupdot \cdots \cupdot \mathcal{Z}_M\) of
the sample space of \(Z\) into \((d+1)\)-dimensional cubes,
\begin{align*}
  \cupdot_{m=1}^{M}\mathcal{Z}_m =  [0, \kappa]; 
\end{align*}
let \(z_m \in \mathcal{Z}_m\) be the midpoint of the cube
\(\mathcal{Z}_m\). We introduce the indicator basis functions
\(\phi_{r}(t)= \1 \lbrace t_{r} \le t \rbrace\) and
\(\phi_{\mathcal{S},m} (z ) = \1 \lbrace z_{m,\mathcal{S}} \le
z_{\mathcal{S}} \rbrace\) which are central components for the
following. Indeed, the discrete approximation \( f_{\beta}\) of \(f\)
with support over these points admits a representation as follows
\begin{align}
  f_{\beta}(t,z) &=  \sum_{r=0}^{R} \phi_{r}(t) \beta_{r}  + \sum_{r=0}^{R}
                   \sum_{\mathcal{S}\subset \lbrace 1,\ldots, d+1\rbrace}
                   \sum_{m =1}^M  \phi_{r}(t)\phi_{\mathcal{S},m} (z) \beta_{r,\mathcal{S},m}. 
  \label{eq:hal:approx:h:11} 
\end{align}
Here, we have that
\(\beta_{r,\mathcal{S},m} = f_{\mathcal{S}} (dt_r, dz_{m,\mathcal{S}})
\), for each
\(r\in \lbrace 1, \ldots, R\rbrace, \mathcal{S}\subset \lbrace
1,\ldots, d+1\rbrace, m=1, \ldots,M\), is the point-mass that the
\(\mathcal{S}\)-specific section of \(f_{\beta} \) assigns to the
point defined by \(z_m\) and \(t_{r-1}\).  Furthermore,
\(\beta_{r} = f_{\beta}(dt_r, 0)\), for \(r=1, \ldots, R\), are the
increments along the time axis alone, and
\(\beta_{0, \mathcal{S},m} = f_{\beta}(0,dz_{m,\mathcal{S}})\), for
\( \mathcal{S}\subset \lbrace 1,\ldots, d+1\rbrace, m = 1, \ldots,
M\), the increments along the \(z\)-axis alone.  Lastly, \(\beta_{0} = f_{\beta}(0,0)\) is the point-mass
assigned by \(f_{\beta}\) to zero. We refer to the stacked vector of
\(\beta_r, \beta_{r,\mathcal{S},m} \) as the vector of parameter
coefficients and note that this vector completely characterizes the
behavior of \(f_{\beta}\). Particularly, the sectional variation norm
of \( f_{\beta}\) becomes a sum over the absolute values of its
coefficients
\begin{align*}
  \Vert f_{\beta} \Vert_v =
  \sum_{r=0}^{R} \vert \beta_{r}\vert  +
  \sum_{r=0}^{R}  \sum_{\mathcal{S}\subset \lbrace 1,\ldots, d+1\rbrace}
  \sum_{m=1}^M  \vert \beta_{r,\mathcal{S},m}\vert
  = \Vert \beta \Vert_1 , 
\end{align*}
i.e., the sectional variation norm of \(f_{\beta}\) equals to the
\(L_1\)-norm of the coefficient vector.  We may now define the highly
adaptive lasso estimator as follows.

\begin{definition}[Highly adaptive lasso estimator]
  The highly adaptive lasso estimator for \(f\) is obtained as
  \(\hat{f}_{n}=f_{\hat{\beta}_n}\) where:
\begin{align}
  \hat{\beta}_n = \underset{ \beta}{\argmin} \,\, \mathbb{P}_n \mathscr{L}(f_\beta),
  \qquad \text{s.t.,} \quad  \Vert \beta \Vert_1  \le \mathscr{M}. 
  \label{eq:hal:lasso:minimization:11}
\end{align}
Notably, \eqref{eq:hal:lasso:minimization:11} corresponds to an
\(L_1\)-penalized regression with indicator functions \(\phi_{r}(t)\)
and \(\phi_{r}(t)\phi_{\mathcal{S},m} (z) \) as covariates and
\(\beta_{r}, \beta_{r,\mathcal{S},m}\) as corresponding coefficients.
\label{def:hal}
\end{definition}

Choosing the support of \(f_\beta\) fine enough, the \(L_1\)-norm of
the coefficient vector approximates the sectional variation norm of
\(f\) and the solution \(f_{\hat{\beta}_n}\) to the minimization
problem defined by \eqref{eq:hal:lasso:minimization:11} in Definition
\ref{def:hal} approximates the infinite-dimensional minimization
problem in \eqref{eq:definition:hal:ideal} over all
\(f \in \mathscr{F}_{\!\! \mathscr{M}} \) \citep[][Appendix
D]{van2017generally}.

We consider particularly the log-likelihood loss function for our
purposes, i.e., we define
\(\mathscr{L}(f)(O) = - \ell_{\mathrm{loglik}}(f)(O)\) where
\((O, f) \mapsto \ell_{\mathrm{loglik}}(f) (O)\) denotes the
log-likelihood
\begin{align*}
  \ell_{\mathrm{loglik}}(f) (O) = \int_0^{\tmax}f(t,Z)\, N(dt) -
  \int_0^{\tmax} \1\lbrace \tilde{T} \ge t\rbrace \exp(f(t,Z)) dt .
\end{align*}
As we review in Appendix B, one way to solve the minimization problem
in \eqref{eq:hal:lasso:minimization:11} in practice with this loss
functions is by standard \(L_1\)-penalized Poisson regression
software, exploiting the correspondence between
\(\ell_{\mathrm{loglik}}\) and the Poisson log-likelihood
\citep{andersen2012statistical,lindsey1995fitting}.  The highly
adaptive lasso estimator \(\hat{f}_{n}=f_{\hat{\beta}_n}\) can next be
plugged into \eqref{eq:parametrization}, providing an estimator for
the hazard itself.  Note that we use cross-validation to
data-adaptively select the bound on the variation norm; by the oracle
properties of cross-validation \citep{van2003unified,van2006oracle},
we only need at least one of the bounds considered as a candidate in
the library to be larger than the true variation norm.

\section{Simulation study}
\label{sec:simulation:study}

In this section we consider a simulation study of estimation of the
average treatment effect parameter
\( \Psi^{\mathrm{ATE}}_{\tmax} \, : \, \mathcal{M} \rightarrow \R\) from
\eqref{eq:target:parameter:ate} specifically. Accordingly, we work
with the clever weights \( w^{\mathrm{ATE}}_t\) from
\eqref{eq:clever:weights:ate}, and the efficient influence function is
given by:
\begin{align*}
  D_{\tmax}^{\mathrm{ATE}}(P) (O) = \int_0^{\tmax} {w}^{\mathrm{ATE}}_t (O)   h_t (O) \, \big( N(dt) -
  \Lambda(dt \, \vert \, A, L)\big) +
  S(\tmax
  \, \vert \, 1, L)  - S(\tmax
  \, \vert \, 0, L) \\
  - \, \Psi^{\mathrm{ATE}}_{\tmax}(P) .
\end{align*}
Our simulation setting imitates the setting of a randomized trial in
which trial participants are randomized to a treatment and followed
over time until either the event or right-censoring
happens. Particularly, \(L=(L_1,L_2,L_3)\) are baseline covariates and
\(A\in \lbrace 0,1\rbrace\) is the randomized treatment. Altogether,
we consider two different versions of the censoring mechanism:
\begin{align*}
 & \qquad \text{Covariate independent censoring: } &
  \lambda^c(t \mid A, L) &=   \lambda^c_0(t),  \\
&  \qquad \text{Covariate dependent censoring: } &
  \lambda^c(t \mid A, L) &=   \lambda^c_0(t) \exp \big(
-0.8L_3 + 1.2L_1A 
  \big).  \qquad\qquad
\end{align*}
Two of the covariates, \(L_1,L_2\), are uniformly distributed on
\((-1,1)\), whereas \(L_3\) is uniform on \((0,1)\). Baseline hazards,
\(\lambda^c_0,\lambda_0\), correspond to Weibull distributions with
shape parameter \(0.7\) and scale parameter \(1.7\) and are the same
across all simulations.  The hazard of the event distribution is given
by:
\begin{align*}
  \lambda(t \mid A, L) = \lambda_0(t) \exp \big(0.7  \1 \lbrace t < t' \rbrace A -0.225
  \1 \lbrace t \ge t' \rbrace A + 
  1.2L_1^2\big), 
\end{align*}
with changepoint \(t'=0.7\). We focus on the survival difference
beyond \(\tmax= 1.2\) years of follow-up and consider the following
different estimation procedures for comparison: 1) A substitution
estimator based on a Kaplan-Meier estimator for the survival curve in
each treatment arm, 2) a targeted maximum likelihood estimator based
on a misspecified Cox model (including main effects for \(A\) and
\(L_1\)) for initial estimation, and 3) a targeted maximum likelihood
estimator based on a Poisson-based highly adaptive lasso estimator for
initial estimation. A grid of ten time-points was used for the time
axis and eight knot-points for each covariate. The upper bound for the
sectional variation norm was selected with cross-validation.  In the
targeting step, we use a correctly specified Cox model for the hazard
of the censoring distribution. The results are presented in Table
\ref{table:SL:tmle:II}, showing that the targeted maximum likelihood
estimator based on the flexible Poisson-based highly adaptive lasso
initial estimation improves precision compared to both the
misspecified Cox model and the Kaplan-Meier
estimator.
 
\begin{table}[ht]
  \centering
  \caption{Results from the simulation study. `HAL-TMLE' uses the
    Poisson-based highly adaptive lasso estimator for initial
    estimation in the targeted maximum likelihood estimation
    algorithm. `Cox-TMLE' uses a misspecified Cox model for initial
    estimation in the targeted maximum likelihood estimation
    algorithm. `KM' uses a Kaplan-Meier approach.  }
\label{table:SL:tmle:II}
\begin{tabular}{rccccccc}
  & \multicolumn{3}{c}{Covariate dependent censoring}
  &
  &  \multicolumn{3}{c}{Covariate independent censoring} \\[0.2em]  
  & HAL-TMLE & Cox-TMLE  & KM & & HAL-TMLE & Cox-TMLE  & KM \\[0.2em]
  \\[-0.5em]
  Bias & -0.0001 & -0.0005 & -0.0059 && 0.0000 & 0.0001 & 0.0002 \\ 
  Cov (95\%) & 0.9540 & 0.9500 & 0.9320 && 0.9360 & 0.9320 & 0.9300 \\ 
  $\sqrt{\mathrm{MSE}}$ & 0.0315 & 0.0321 & 0.0326 && 0.0310 & 0.0316 & 0.0316 \\ 
  rel. MSE & 0.9344 & 0.9703 & 1.0000 && 0.9600 & 0.9988 & 1.0000 \\[-0.75em]    
\end{tabular}
\end{table}

\section{Concluding remarks}
\label{sec:discussion}

Our simulations show that the targeted maximum likelihood estimator
based on the implemented Poisson-based highly adaptive lasso estimator
performs in agreement with the asymptotic theory, improving precision
relative to the Kaplan-Meier approach and achieving proper coverage
based on the efficient influence function.

To fully optimize the estimation of all nuisance parameters, which are
comprised by the hazard for the censoring and event distributions and
the conditional distribution of treatment, we recommend to apply
loss-function based cross-validation to combine the highly adaptive
lasso estimator with other estimators.  This procedure of selecting
the best estimator from of a prespecified library of candidate
algorithms by minimizing the cross-validated empirical risk is often
referred to as super learning \citep{van2007super} and the general
oracle inequality for loss-based cross-validation
\citep{van2003unified,van2006oracle} yields that the super learner
will achieve the minimal rate of convergence of the estimators in the
library.  In future work, we plan to establish the oracle inequality
for the considered loss function, involving understanding that a
discrete baseline hazard is really the appropriate approach (see also
Appendix B).\color{black}

\section*{Supplementary material}

Supplementary material includes I) a sketch of the proof of causal
interpretability; II) the proof for the sufficiency of the conditions
stated in Section \ref{sec:conditions:for:al:eff}; III) derivations of
efficient influence functions and of second\hyp{}order remainders for
the survival analysis setting; IV) derivations of efficient influence
functions and of second\hyp{}order remainders for the competing risks
setting; V) descriptions of implementations; and VI) additional
simulations for the competing risks setting.

\appendix

\section*{Appendix A}

\renewcommand\theequation{A.\arabic{equation}}
\setcounter{equation}{0}

\subsection*{Targeted maximum likelihood estimation for the competing
  risks setting}

We here consider the competing risks analogue of the observed data
setting considered in the main text. Let \(A \in \lbrace 0,1\rbrace\),
\(L \in \R^d\) and the time under observation \(\tilde{T}\in \R_+\) be
as in Section \ref{sec:actual:setting:notation}. The variables
\(T\in \R_+\) and \(C\in \R_+\) represent the times to event, now one
of \(J\ge 1\) types, and censoring, respectively, such that the
observed time is \(\tilde{T} = \min ( T, C)\). Together,
\(O=(L, A, \tilde{T}, \tilde{\Delta})\) constitutes the observed data.
We further define an event indicator
\(\Delta \in \lbrace 1,2, \ldots, J\rbrace\) telling us the type of
event happening. Note that \(\Delta\) is subject to right-censoring
such that we only observe
\(\tilde{\Delta} = \1\lbrace T \ge C \rbrace \Delta\). We define the
counting processes
\(N(t) = (N_1(t), \ldots, N_J(t)) = (\1\lbrace \tilde{T} \le t,
\tilde{\Delta}=1\rbrace, \ldots,\1\lbrace \tilde{T} \le t,
\tilde{\Delta}=J\rbrace) \) and
\(N^c(t) = \1\lbrace \tilde{T} \le t, \Delta=0\rbrace\). Let
\(\lambda_{0,j}\) denote the cause \(j\) specific hazard, defined as
\begin{align*}
  \lambda_{0,j}(t \, \vert \, a, \ell)
  & =
    \underset{h \rightarrow 0}{\lim} \,\, h^{-1} P_0(T \le t+h, \Delta= j \mid \tilde{T} \ge t,
     A=a, L=\ell),
\end{align*}
and \(\Lambda_{0,j}\) the corresponding cumulative hazard.
The interventional part of the likelihood is unchanged, whereas the
non-interventional part is given by:
\begin{align*}
  q_{0} (o)
  & =
    \mu_0(\ell) 
    \prod_{j=1}^J  \big( \lambda_{0,j} (t \mid a, \ell) \big)^{\1 \lbrace \delta = j\rbrace}
    S_0(t- \mid a, \ell), 
\end{align*}
where
\(S_0 (t \mid a, \ell) = \exp (- \int_0^t \sum_{j=1}^J
\lambda_{0,j}(s \mid a, \ell) ds) \).  With no loss of generality,
we assume that \(J=2\). We define the target parameter
\(\Psi_{\tmax} \, : \, \mathcal{M} \rightarrow \R\) as the
intervention-specific absolute risk beyond time \(\tmax\):
\begin{align}
  \begin{split}
    \Psi_{\tmax}(P) & = \int_{\mathcal{L}}\sum_{a=0,1} F_1(\tmax \mid
    a, \ell) \pi^*(a\mid L) \mu ( \ell)d\nu(\ell) ,
  \end{split}
  \label{eq:CR:target:parameter}
\end{align}
where
\begin{align*}
  F_1(t \mid a,\ell)
  = \int_0^t S(s- \mid a, \ell) \Lambda_1 (ds \mid a, \ell) , 
\end{align*}
is the risk function for the event of interest (\(j=1\)) 
\citep{gray1988class}. We define clever weights and covariates
according to Definitions \ref{defi:clever:weigths:cr} and
\ref{defi:clever:covariates:cr} as follows.

\begin{definition}[Clever weights]
\label{defi:clever:weigths:cr}
Define clever weights by:
\begin{align*}
  w_t(O) & =
\1 \lbrace \tilde{T} \ge t \rbrace\frac{
      \pi^*(A \mid L) }{ \pi (A \, \vert \, L) } \frac{\1 \lbrace t  \le \tmax \rbrace}{
      {S}^c( t- \, \vert \, A,L)
    }.
\end{align*}
\end{definition}

The clever weights are the same as for the survival analysis
setting (Definition \ref{defi:clever:weigths}). 

\begin{definition}[Clever covariates]
  \label{defi:clever:covariates:cr}
  Define clever covariates by:
  \begin{align*}
    h_{1,t} (O) & = 
                  1- \frac{ {F}_1 ( \tmax  \mid A, L) - {F}_1 ( t \mid A, L)}{
                  {S}(t \, \vert \, A,L)}   \\
    h_{2,t} (O) & =
                  -\frac{ {F}_1 ( \tmax  \mid A, L) - {F}_1 ( t \mid A, L)}{
                  {S}(t \, \vert \, A,L) } ,
  \end{align*}
  for \(t>0\). 
\end{definition}

Now we can define a targeted maximum likelihood algorithm for
estimation of the target parameter \eqref{eq:CR:target:parameter}. The
algorithm involves targeting steps for both the intensity of the event
process of interest \(\lambda_1\) and then intensity of the competing
event process \(\lambda_2\). We define the fluctuation models as
follows:
\begin{align}
  \lambda_1 (t; \eps_1) & = \lambda_1(t) \exp ( \eps_1 {w}_t {h}_{1,t}), \qquad \eps_1\in\R, 
                          \label{eq:lambda:1:fluc} \\
  \lambda_2 (t; \eps_2) & = \lambda_2(t) \exp ( \eps_2 {w}_t {h}_{2,t}), \qquad \eps_2\in\R.
  \label{eq:lambda:2:fluc}
\end{align}

As the setting of Section \ref{sec:tmle}, \(\hat{\pi}_n\) and
\(\hat{\lambda}^c_n\) together define an estimator \( \hat{w}_{t}\)
for the clever weights, but now estimators of the clever covariates
\(\hat{h}_{1,t}\), \(\hat{h}_{2,t}\) need both \(\hat{\lambda}_{1,n}\)
and \(\hat{\lambda}_{2,n}\). In each round of iterations, the
evaluation of \eqref{eq:lambda:1:fluc} in the current estimator
\(\hat{\lambda}_{1,n,k}\) and the evaluation of
\eqref{eq:lambda:2:fluc} in the current estimator
\(\hat{\lambda}_{2,n,k}\) together with the estimator \( \hat{w}_{t}\)
for the clever weights defines one equation in \(\eps_1\) and one in
\(\eps_2\): 
\begin{align}
  \begin{split}
  &0= \frac{1}{n} \sum_{i=1}^n \bigg( 
  \int_0^{\tmax}  \hat{w}_t (O_i) \hat{h}_{j,t,k} (O_i)\, N_{j,i}(dt)  \\
  &\qquad\qquad - \int_0^{\tmax} \hat{w}_t (O_i)\hat{h}_{j,t,k} (O_i)
    \exp (\eps_j \hat{w}_t (O_i) \hat{h}_{j,t,k} (O_i) ) \, \hat{\lambda}_{j,n,k}
  (t \mid A_i, L_i ) dt \bigg) 
, \quad j=1,2. 
    \end{split}\label{eq:equation:written:out:0}
\end{align}
The solution \(\hat{\eps}_{j,k}\), \(j=1,2\), defines the update of
\(\hat{\lambda}_{j, n,k}\), \(j=1,2\), along the corresponding
fluctuation model:
\begin{align*}
  \hat{\lambda}_{j,n,k+1} (t) := \hat{\lambda}_{j,n,k}(t) \exp ( \hat{\eps}_{j,k} \hat{w}_t \hat{h}_{j,t,k}),
  \qquad j=1,2,\quad
  k \ge 0 . 
\end{align*}
The steps from \(k\) to \(k+1\) are repeated until convergence.

\section*{Appendix B}

\renewcommand\theequation{B.\arabic{equation}}
\setcounter{equation}{0}

In this appendix, we sketch a result for the highly adaptive lasso
estimator presented in Section \ref{sec:poisson:hal} of the main text:
That we can implement the highly adaptive lasso estimator (Definition
\ref{def:hal}) in practice with \(L_1\)-penalized Poisson regression
software.

The fact that the likelihood for a proportional hazards model
corresponds to that of a certain Poisson regression is a well-known
result from the survival analysis literature
\citep{andersen2012statistical,lindsey1995fitting}. The technique is
commonly applied in large-scale observational studies
\citep{lin1998assessing,gron2016misspecified} to approximate Cox
regression models in a way that can potentially save considerable
computation time and memory usage.  Put shortly, the Poisson
formulation is just a different formulation of a proportional hazards
model with a baseline rate modeled by a parameter over a grid of
time-points, assuming a constant rate in each interval between
time-points times which allows for the use of standard Poisson
regression software. We emphasize that the Poisson regression is only
used as a tool; never is the Poisson model in fact assumed for the
data. Furthermore, although the formulation is over a discrete
time-points, the exact continuous event times are still used in the
estimation, specifically in the aggregated risk time used as an offset
in the regression.

\subsection*{Highly adaptive lasso estimation implemented as a Poisson regression}
\label{app:why:poisson}

We here clarify how one may in practice solve the minimization problem
of Definition \ref{def:hal} for the highly adaptive lasso estimator by
standard \(L_1\)-penalized Poisson regression software.  The loss
function used in Definition \ref{def:hal} to define the highly
adaptive lasso for a hazard on the form
\( \lambda (t \, \vert \, Z) = \exp ( f(t, Z) )\) is the
log-likelihood loss function
\(\mathscr{L}(f)(O) = - \ell_{\mathrm{loglik}}(f)(O)\) where
\((O, f) \mapsto \ell_{\mathrm{loglik}}(f) (O)\) is given by
\begin{align*}
  \ell_{\mathrm{loglik}}(f) (O) = \int_0^{\tmax}f(t,Z)\, N(dt) -
  \int_0^{\tmax} \1\lbrace \tilde{T} \ge t\rbrace \exp(f(t,Z)) dt.
\end{align*}
Particularly, for \(f_\beta\) admitting the representation in
\eqref{eq:hal:approx:h:11},
\begin{align*}
  f_{\beta}(t,z) &=  \sum_{r=0}^{R} \phi_{r}(t) \beta_{r}  + \sum_{r=0}^{R}
                   \sum_{\mathcal{S}\subset \lbrace 1,\ldots, d+1\rbrace}
                   \sum_{j \in \mathcal{I}_{\mathcal{S}}}  \phi_{r}(t)\phi_{\mathcal{S},j} (z) \beta_{r,\mathcal{S},j},
\end{align*}
we have that
\begin{align}
  \begin{split}
    & \ell_{\mathrm{loglik}}(f_\beta) (O) = \sum_{r=0}^R \big( N(t_{r+1} -
    N(t_r) \big)
    f_{\beta} (t_{r} , Z)  \\[-0.3cm]
    &\qquad\qquad\qquad\qquad - \, \sum_{r=0}^{R} \1\lbrace \tilde{T}
    \ge t_{r} \rbrace \exp ( f_{\beta} (t_{r} , Z) ) \big(\min
    (\tilde{T}, t_{r+1} )-t_{r}\big).
\end{split}
  \label{eq:log:like:2}
\end{align}
Applying our notation from Section \ref{sec:poisson:hal} of the main
text with a partitioning
\(\mathcal{Z}_1 \cupdot \cdots \cupdot \mathcal{Z}_M\) of the sample
space of \(Z\) into \((d+1)\)-dimensional cubes, \(m\) is an index
that runs through distinct values of the vector
\(( \phi_{\mathcal{S},m} (Z) \, :\, \mathcal{S},m)\), i.e., for each
\(m=1,\ldots,M\),
\begin{align*}
  \forall \,z^1,z^2 \in  \mathcal{Z}_m : \quad
  \phi_{\mathcal{S},m} (z^1)   =  \phi_{\mathcal{S},m} (z^2),  \quad \text{for all } \mathcal{S},m.
\end{align*}
Let again \(z_m \in \mathcal{Z}_m\) be the midpoint of the cube
\(\mathcal{Z}_m\). Now, observe on the one hand that
\begin{align*}
  &  \sum_{i=1}^n  \sum_{r=0}^{R} \big(N_i (t_{r+1}) - N_i(t_{r})\big) f_\beta(t_{r}, Z_i) \\[-0.5em]
  &\qquad\qquad\qquad = \sum_{m=1}^M \sum_{r=0}^R
    \underbrace{\sum_{i=1}^n \big(N_i (t_{r+1}) - N_i(t_{r})\big)
    \1 \lbrace Z_i \in \mathcal{Z}_m \rbrace }_{=:\mathcal{D}_{m,r}}{f}_{\beta}(t_{r}, z(m)) ,
\end{align*}
and, on the other hand, that
\begin{align*}
  &\sum_{i=1}^n  \sum_{r=0}^{R}    \1\lbrace \tilde{T}_i \ge t_{r}\rbrace   \big( \min (\tilde{T}_i, t_{r+1})
    -
    t_{r}\big)
    \exp ({f}_{\beta}( t_{r}, Z_i) ) \\[-0.5em]
  &\qquad\qquad =
    \sum_{m=1}^M    \sum_{r=0}^{R}  \underbrace{\sum_{i=1}^n
    \1\lbrace \tilde{T}_i \ge t_{r}\rbrace   \big( \min (\tilde{T}_i, t_{r+1})
    -
    t_{r}\big) \1 \lbrace Z_i \in \mathcal{Z}_m \rbrace }_{=:\mathcal{R}_{m,r}}
    \exp ({f}_{\beta}( t_{r}, z(m)) )  ,
\end{align*}
from which we see that
\begin{align}
  &  \underset{ \beta}{\textnormal{\argmin}} \, - \mathbb{P}_n \ell_{\mathrm{loglik}}(f_\beta)\notag \\
  &\qquad =  \underset{ \beta}{\textnormal{\argmin}} \, - \bigg( \sum_{i=1}^n \sum_{r=0}^R \big( N(t_{r+1} -
    N(t_r) \big) f_\beta(t_{r}, Z_i)
    \notag \\[-0.95em]
  &\qquad\qquad\qquad\qquad\qquad\qquad - \,  
    \sum_{r=0}^{R}
    \1\lbrace \tilde{T}_i \ge t_{r}\rbrace \big( \min (\tilde{T}_i, t_{r+1})
    -
    t_{r}\big)
    \exp ( {f}_{\beta}( t_{r}, Z_i) )  \color{black} \bigg)  \notag\\[-0.4em]
  & \qquad =  \underset{ \beta}{\textnormal{\argmin}} \, - \sum_{m=1}^M \sum_{r=0}^{R}
    \Big(  \mathcal{D}_{m,r} {f}_{\beta}(t_{r}, z_m) -
    \mathcal{R}_{m,r}  \exp ( {f}_{\beta}(t_{r},z_m) ) \Big). \label{eq:losss:1}
\end{align}
Importantly, \eqref{eq:losss:1} corresponds to minimizing a Poisson
likelihood loss function with event counts \(\mathcal{D}_{m,r} \),
mean \( \exp( {f}_{\beta}(t_r, z_m))\) and offset
\(\log \mathcal{R}_{m,r}\). So, in practice we can fit this Poisson
regression model, working with the aggregated data (where, for each
combination of covariates values, \(z_1, \ldots, z_M\), and
intervals \([ t_{r},t_{r+1})\), we count events and we sum up the
total risk time).

\subsection*{Convergence of the highly adaptive lasso estimator
  implemented as a Poisson regression}
\label{app:poisson:convergence}

For the log-likelihood loss function
\(\mathscr{L}(f)(O) = - \ell_{\mathrm{loglik}}(f)(O)\), define as
follows
\begin{align}
  \hat{f}^*_n = \underset{ f \in   \mathscr{F}_{\!\!\mathscr{M}}  }{\argmin} \,\,
  \mathbb{P}_n \mathscr{L}(f).
  \label{eq:hal:lasso:minimization:21}
\end{align}
For a constant \(\mathscr{M}'<\infty\) we make the assumptions that
\begin{align}
  \sup_{f\in \mathscr{F}_{\!\!\mathscr{M}}}  \frac{P_0
  \big( \mathscr{L} (f )
  -  \mathscr{L} (f_0 ) \big)^2}{
  P_0 \big( \mathscr{L} (f )
  -  \mathscr{L} (f_0 ) \big)
  }\le  \mathscr{M}'
  ,\label{eq:bound:ass}
\end{align}
and, 
\begin{align}
  \sup_{f\in \mathscr{F}_{\!\!\mathscr{M}}}  \frac{
  \Vert \mathscr{L}(f) \Vert_v}{
  \Vert f \Vert_v  } <\infty \label{eq:bound:ass:2}. 
\end{align}
\color{black} The following arguments follow previous work
\citep[][Chapters 6,7]{van2017generally,van2018targeted}, but are
repeated here for completeness.  Consider the following bound for the
log-likelihood based dissimilarity:
\begin{align}
      \begin{split}
        &P_0 \big( \mathscr{L} (\hat{f}^*_n )
        -  \mathscr{L} (f_0 ) \big) \\
        &\qquad = - (\mathbb{P}_n - P_0) \big( \mathscr{L}
        (\hat{f}^*_n ) - \mathscr{L} (f_0 ) \big) + \mathbb{P}_n \big(
        \mathscr{L} (\hat{f}^*_n )
        -  \mathscr{L} (f_0 ) \big) \\
        &\qquad \le - (\mathbb{P}_n - P_0) \big( \mathscr{L}
        (\hat{f}^*_n ) - \mathscr{L} (f_0 ) \big),
    \end{split}\label{eq:bound:loss:based:diss}
\end{align}
where, at the inequality, we used
\eqref{eq:hal:lasso:minimization:21}. It is a straightforward
consequence of Assumption \ref{ass:overall:cadlag} and the bound from
\eqref{eq:bound:ass:2} that
\( \mathscr{L} (\hat{f}^*_n ) - \mathscr{L} (f_0
) \) belongs to a Donsker class. Thus, we have that
\begin{align*}
 - (\mathbb{P}_n - P_0) \big( \mathscr{L} (\hat{f}^*_n )
    -  \mathscr{L} (f_0 ) \big) = O_P(n^{-1/2}),
\end{align*}
which, combined with \eqref{eq:bound:loss:based:diss} and the bound
from \eqref{eq:bound:ass} implies that
\begin{align*}
(\mathbb{P}_n - P_0) \big( \mathscr{L} (\hat{f}^*_n )
    -  \mathscr{L} (f_0 ) \big) = o_P(n^{-1/2}), 
\end{align*}
by \citet[][Lemma 19.24]{van2000asymptotic}. Thus, again by
\eqref{eq:bound:loss:based:diss}, it follows that
\begin{align}
  P_0 \big( \mathscr{L} (\hat{f}^*_n )
  -  \mathscr{L} (f_0 ) \big) = o_P(n^{-1/2}).
\label{eq:convergence:loss:based:diss}
\end{align}

We next show that the highly adaptive lasso estimator from Definition
\ref{def:hal} fulfills the same convergence condition for a
partitioning that is chosen fine enough. For the following we let
\(\gridsize >0 \) denote the maximal side length of each cube of the
partitioning for a given partitioning of \([0, \tmax]\) and
\([0,\kappa]\). Then, for given \(\gridsize\) and a constant
\(\mathscr{M}<\infty\), we define as follows:
\begin{align}
  \begin{split}
    & \mathscr{F}^{\gridsize}_{\!\!\mathscr{M}} = \Bigg\lbrace
    \sum_{r=0}^{R} \phi_{r}(t) \beta_{r} + \sum_{r=0}^{R}
    \sum_{\mathcal{S}\subset \lbrace 1,\ldots, d+1\rbrace}
    \sum_{m=1}^M \phi_{r}(t)\phi_{\mathcal{S},m} (z)
    \beta_{r,\mathcal{S},m}
    \, :\\
    &\qquad\qquad\qquad\qquad\qquad\qquad \Vert \beta \Vert_1 =
    \sum_{r=0}^{R} \vert \beta_{r}\vert + \sum_{r=0}^{R}
    \sum_{\mathcal{S}\subset \lbrace 1,\ldots, d+1\rbrace}
    \sum_{m=1}^M \vert \beta_{r,\mathcal{S},m}\vert \le \mathscr{M}
    \Bigg\rbrace.
      \end{split}\label{eq:defi:F:beta}
\end{align}
Now the highly adaptive lasso estimator from Definition \ref{def:hal}
can be written as follows
\begin{align}
  \hat{f}^{\gridsize}_n = \underset{ f \in   \mathscr{F}^{\gridsize}_{\!\!\mathscr{M}}  }{\argmin} \,\,
  \mathbb{P}_n \mathscr{L}(f)
  . 
  \label{eq:hal:lasso:minimization:22}
\end{align}
Consider
\begin{align*}
      \begin{split}
        &P_0 \big( \mathscr{L} (\hat{f}^{\gridsize}_n )
        -  \mathscr{L} (f_0 ) \big) \\
        &\qquad = - (\mathbb{P}_n - P_0) \big( \mathscr{L}
        (\hat{f}^{\gridsize}_n ) - \mathscr{L} (f_0 ) \big) +
        \mathbb{P}_n \big( \mathscr{L}
        (\hat{f}^{\gridsize}_n ) - \mathscr{L} (f_0 ) \big) \\
        &\qquad = - (\mathbb{P}_n - P_0) \big( \mathscr{L}
        (\hat{f}^{\gridsize}_n ) - \mathscr{L} (f_0 ) \big) +
        \mathbb{P}_n \big( \mathscr{L}
        (\hat{f}^{\gridsize}_n ) - \mathscr{L} (\hat{f}^*_n ) \big) +
        \mathbb{P}_n \big( \mathscr{L}
        (\hat{f}^*_n ) - \mathscr{L} (f_0 ) \big) \\
        &\qquad \le - (\mathbb{P}_n - P_0) \big( \mathscr{L}
        (\hat{f}^{\gridsize}_n ) - \mathscr{L} (f_0 ) \big) +  \mathbb{P}_n \big( \mathscr{L}
        (\hat{f}^{\gridsize}_n ) - \mathscr{L} (\hat{f}^*_n ) \big).
    \end{split}
\end{align*}
Following the arguments from \eqref{eq:bound:loss:based:diss} to
\eqref{eq:convergence:loss:based:diss} above with
\(\hat{f}^{\gridsize}_n\) substituted for \(\hat{f}^*_n\) yields that
\begin{align*}
(\mathbb{P}_n - P_0) \big( \mathscr{L} (\hat{f}^{\gridsize}_n ) -
  \mathscr{L} (f_0 ) \big) = o_P(n^{-1/2}).
\end{align*}
Furthermore, \citet[][Lemma 11]{van2017generally} gives that
\begin{align*}
  \mathbb{P}_n \big( \mathscr{L} ( \hat{f}^{\gridsize}_n  )
  -  \mathscr{L} (\hat{f}^*_n ) \big)
  \rightarrow 0 ,  \qquad \text{as} \quad \gridsize \rightarrow 0;
\end{align*}
particularly, if \(m = o(n^{-1/2})\) then
\(\mathbb{P}_n \big( \mathscr{L} ( \hat{f}^{\gridsize}_n ) -
\mathscr{L} (\hat{f}^*_n ) \big) = o_P(n^{-1/2})\) \color{black} and
it follows that
\begin{align*}
 P_0 \big( \mathscr{L} (\hat{f}^{\gridsize}_n )
    -  \mathscr{L} (f_0 ) \big) = o_P(n^{-1/2}),
\end{align*}
as desired. In Appendix C we outline the arguments to show that
convergence in terms of the loss-based dissimilarity implies the same
convergence rate for the squared
\(L_2(\pi^* \otimes \mu_0\otimes \rho)\); i.e., we show that
\begin{align*}
 P_0 \big( \mathscr{L} (\hat{f}^{\gridsize}_n )
    -  \mathscr{L} (f_0 ) \big) = o_P(n^{-1/2}),
\end{align*}
implies for
\(\lambda_{f^{\gridsize}} (t\mid Z) = \exp(f^{\gridsize}(t, Z))\) that
\begin{align*}
  \Vert \lambda_{f^{\gridsize}} - \lambda_0 \Vert_{\pi^* \otimes \mu_0\otimes \rho}^2 =
  o_P(n^{-1/2}), 
\end{align*}
which was exactly what we needed to control the second-order
remainder, see Section \ref{remark:R2} of the main text.

\section*{Appendix C}

\subsection*{Rate of convergence}

It is a general result that the likelihood loss-based dissimilarity,
which is really the Kullback-Leibler dissimilarity, behaves as a
square of an \(L_2(P_0 )\)-norm \citep[see, e.g.][Lemma
4]{van2017generally} when the density \(p_0\) of the distribution
\(P_0\) is bounded. We here repeat the arguments to demonstrate that
the highly adaptive lasso convergence in terms of the loss-based
dissimilarity also implies the needed convergence to control the
second-order remainder (Section \ref{remark:R2}).

The general arguments follow \citet[][p. 62]{van2000asymptotic} for
the density \(p_0\) of the observed data distribution \(P_0\).
Indeed, the Kullback-Leibler dissimilarity for a \(P\in \mathcal{M}\)
with density \(p\) with respect to the same dominating measure \(\nu\)
can be bounded as follows:
\begin{align}
  \begin{split}
    & \int \log \bigg( \frac{p}{p_0}\bigg) (o) dP_0(o)
    \intertext{here, we use that \(\log x \le 2 (\sqrt{x} - 1)\) for
      \(x\ge 0\);} & \qquad \le 2 \int \bigg(
    \frac{\sqrt{p}}{\sqrt{p_0}} -1\bigg) (o)
    dP_0 (o) \\
    & \qquad = 2 \int \bigg(\frac{\sqrt{p}}{\sqrt{p_0}}\bigg) (o)
    p_0 (o) d\nu(o) - 2 \\
    & \qquad = 2 \int \big( \sqrt{p}\sqrt{p_0}\big) (o) d\nu(o) - 2
    \notag
  \end{split} \\
    & \qquad \le - \int \big( \sqrt{p} - \sqrt{p} \big)^2 (o) d\nu(o)
      , \label{eq:last:112}
\end{align}
here, for the last inequality we used that
\begin{align*}
  & \int \big( \sqrt{p} -
    \sqrt{p_{0}} \big)^2 (o)  d\nu(o)  \\
  &\qquad = \int p(o)  d\nu(o)  +
    \int p_{0}(o)  d\nu(o)  - 2
    \int \big( \sqrt{pp_{0}} \big)(o)  d\nu(o)  \\
  &\qquad \le 2 - 2  \int \big( \sqrt{pp_{0}} \big) (o)  d\nu(o)  . 
\end{align*}
On the other hand we have that
\begin{align}
  \begin{split}
    & \int \big( p - p_0 \big)^2 (o) d\nu(o) = \int \big( \sqrt{p} -
    \sqrt{p_0} \big)^2 \underbrace{\big( \sqrt{p} + \sqrt{p_0}
      \big)^2}_{\le \mathscr{M}'} (o) d\nu(o)\\
    & \qquad \le \mathscr{M}' \int \big( \sqrt{p} - \sqrt{p_0} \big)^2
    (o) d\nu(o)\le - \mathscr{M}' \int \log \bigg( \frac{p}{p_0}\bigg) (o)
    dP_0(o),
    \end{split}\label{eq:last:113}
\end{align}
using \eqref{eq:last:112} and that the density is bounded. We see that
\eqref{eq:last:113} confirms the general claim. \\

Next, consider our observed data distribution \(P_0\) of
\(O= (L, A, \tilde{T}, \Delta)\).  For the following, we consider the
conditional distribution of \((\tilde{T},\Delta)\) given \(A, L\)
which we shall denote by \(\tilde{P}_0\) with density \(\tilde{p}_0\),
i.e.,
\begin{align*}
  \tilde{p}_0(o) = \underbrace{ \big(
  {\lambda}_0 (t \mid a, \ell)\big)^{\delta}  {S}_0(t  \mid a,
  \ell)}_{=:\tilde{q}_0 (t, \delta \mid a, \ell)}
  \underbrace{ \big( {\lambda}^c_0 (t \mid a, \ell)
  \big)^{1-\delta} {S}_0^c(t  \mid a, \ell)}_{=:\tilde{g}_0 (t, \delta \mid a, \ell)} = \tilde{q}_0 (t, \delta \mid a, \ell)
  \tilde{g}_0 (t, \delta \mid a, \ell). 
\end{align*}
For the following, we further denote by
\begin{align*}
  p_{\lambda}(t  \mid a, \ell)= 
  {\lambda} (t \mid a, \ell){S}(t  \mid a,
  \ell)
  , 
\end{align*}
the conditional density of the distribution of \(T\). Note that
\(p_{\lambda} ( t\mid a ,\ell) = \tilde{q} (t , 1\mid a, \ell)\).

Repeating the arguments of \eqref{eq:last:112} above conditional on
fixed \(a,\ell\) yields the following bound in terms of the
Kullback-Leibler dissimilarity
\begin{align*}
  \sum_{\delta =0,1}\int_0^{\tmax} \big( \sqrt{\tilde{p}} - \sqrt{\tilde{p}}_0 \big)^2   (t, \delta \mid a,\ell) dt
  \le
  - \sum_{\delta =0,1}\int_0^{\tmax} \log \bigg( \frac{\tilde{p}}{\tilde{p}_0}\bigg) (t, \delta \mid a,\ell)
  \tilde{p}_0 (t, \delta \mid a, \ell)dt . 
\end{align*}
Particularly, note that we only care about the \(\tilde{q}_0\)-factor;
due to the factorization of \(\tilde{p}_0\) displayed above, we can
act as if \(\tilde{g}_0\) is known, i.e.,
\(\tilde{p} = \tilde{q} \tilde{g}_0\) as well as
\(\tilde{p}_0 = \tilde{q}_0 \tilde{g}_0\). For the left hand side of
the above, we thus have that
\begin{align*}
  & - \sum_{\delta =0,1}\int_0^{\tmax} \big( \sqrt{\tilde{p}} -
    \sqrt{\tilde{p}_{0}} \big)^2 (t, \delta \mid a,\ell) dt
    = - \sum_{\delta =0,1}\int_0^{\tmax} \big( \sqrt{\tilde{q}} -
    \sqrt{\tilde{q}_{0}} \big)^2(t, \delta \mid a,\ell) \tilde{g}_{0}(t, \delta \mid a,\ell) dt \\
  & \qquad  \overset{*}{\le} - \tilde{\eta} \sum_{\delta =0,1}\int_0^{\tmax} \big( \sqrt{\tilde{q}} -
    \sqrt{\tilde{q}_{0}} \big)^2(t, \delta \mid a,\ell)  dt , 
\end{align*}
where \(*\) follows under the assumption that \(\tilde{g}_0\) is
bounded away from zero for all \(a,\ell\) by \(\tilde{\eta}>0\).  Now
we see that
\begin{align*}
  & \sum_{\delta =0,1}\int_0^{\tmax} \big( \tilde{q} -
    \tilde{q}_0 \big)^2 (t, \delta \mid a,\ell) dt
    =  \sum_{\delta =0,1}\int_0^{\tmax} \big( \sqrt{\tilde{q}} -
    \sqrt{\tilde{q}_0} \big)^2  \underbrace{\big( \sqrt{\tilde{q}} +
    \sqrt{\tilde{q}_0} \big)^2}_{\le  \tilde{\mathscr{M}}'} (t, \delta \mid a,\ell) dt  \\[-0.1cm]
  & \qquad \le \tilde{\mathscr{M}}'
    \sum_{\delta =0,1}\int_0^{\tmax} \big( \sqrt{\tilde{q}} -
    \sqrt{\tilde{q}_0} \big)^2   (t, \delta \mid a,\ell) dt \\
  & \qquad 
    \le -\tilde{\mathscr{M}}' \tilde{\eta}^{-1}\sum_{\delta =0,1}\int_0^{\tmax}
    \log \bigg( \frac{\tilde{p}}{\tilde{p}_0}\bigg) (t, \delta \mid a,\ell)
    \tilde{p}_0 (t, \delta \mid a, \ell)dt, 
\end{align*}
i.e., we have that
\begin{align}
  \begin{split}
    & \int_{\mathcal{L}} \sum_{a=0,1} \bigg(\sum_{\delta
      =0,1}\int_0^{\tmax} \big( \tilde{q} -
    \tilde{q}_0 \big)^2 (t, \delta \mid a,\ell) dt \bigg) \pi_0 (a\mid \ell) d\mu_0(\ell) \\
    & \qquad \le \tilde{\mathscr{M}}' \tilde{\eta}^{-1}
    \int_{\mathcal{L}} \sum_{a=0,1} \bigg( \sum_{\delta
      =0,1}\int_0^{\tmax} \log \bigg(
    \frac{\tilde{p}}{\tilde{p}_0}\bigg) (t, \delta \mid a,\ell)
    \tilde{p}_0 (t, \delta \mid a, \ell)dt\bigg) \pi_0 (a\mid \ell)
    d\mu_0(\ell).
  \end{split}\label{eq:last:ineq:114}
\end{align}
By the general result for highly adaptive lasso estimation
\citep{van2017generally} we have the following convergence with
respect to the Kullback-Leibler dissimilarity
\begin{align}
\int_{\mathcal{L}}  \sum_{a=0,1} \bigg(\sum_{\delta =0,1}\int_0^{\tmax}
    \log \bigg( \frac{\tilde{p}}{\tilde{p}_0}\bigg) (t, \delta \mid a,\ell)
  \tilde{p}_0 (t, \delta \mid a, \ell)dt \bigg) \pi_0 (a\mid \ell) d\mu_0(\ell)= o_P(n^{-1/2}),
  \label{eq:hal:results}
\end{align}
particularly, observe that
\begin{align*}
  &\sum_{\delta =0,1}\int_0^{\tmax} \big( \tilde{q} -
    \tilde{q}_0 \big)^2 (t, \delta \mid a,\ell) dt
    =  \int_0^{\tmax} \big( p_{\lambda} -
    p_{\lambda_0} \big)^2 (t \mid a,\ell) dt
    +  \int_0^{\tmax} \big( S -
    S_0\big)^2 (t \mid a,\ell) dt,
\end{align*}
so that combining \eqref{eq:last:ineq:114} and \eqref{eq:hal:results}
yields that
\( \Vert p_{\lambda} - p_{\lambda_0} \Vert_{\pi^* \otimes \mu_0\otimes
  \rho} =o_P(n^{-1/4})\). To realize that this also implies that
\( \Vert \lambda- \lambda_0 \Vert_{\pi^* \otimes \mu_0\otimes \rho} =
o_P(n^{-1/4})\), observe that
\begin{align*}
  &\Vert \lambda- \lambda_0 \Vert_{\pi^* \otimes
    \mu_0\otimes \rho} = \int_{\mathcal{L}} \sum_{a=0,1} \bigg(
    \int_0^{\tmax} \big( \lambda- \lambda_0 \big)^2 (t \mid a, \ell) dt
    \bigg) \pi^*(a \mid \ell) \mu_0(\ell)d\nu(\ell) \intertext{since
    \(p_{\lambda} = \lambda S\), this is the same as}
  &\qquad = \int_{\mathcal{L}}
    \sum_{a=0,1}\bigg( \int_0^{\tmax} \bigg( \frac{p_{\lambda}}{S} -
    \frac{p_{\lambda_0}}{S_0} \bigg)^2 (t \mid a, \ell) dt
    \bigg)    \pi^*(a \mid \ell) \mu_0(\ell)d\nu(\ell)\\
  &\qquad  =  \int_{\mathcal{L}} \sum_{a=0,1} \bigg(  \int_0^{\tmax} \bigg( \frac{p_{\lambda}-p_{\lambda_0}}{S_0} + \frac{p_{\lambda}}{S} - \frac{p_{\lambda}}{S_0}  \bigg)^2 (t \mid a, \ell)  dt    \bigg)    \pi^*(a \mid \ell) \mu_0(\ell)d\nu(\ell)\\
  &\qquad = \int_{\mathcal{L}} \sum_{a=0,1} \bigg(
    \int_0^{\tmax} \bigg(
    \frac{p_{\lambda}-p_{\lambda_0}}{S_0} +
    p_{\lambda}\,\frac{S_0 - S}{S S_0} \bigg)^2
    (t \mid a, \ell) dt \bigg) \pi^*(a \mid \ell)
    \mu_0(\ell)d\nu(\ell) \intertext{here we can
    use that \(S,S_0\) are bounded away from zero
    by some \(\eta',\eta'_0>0\) on the bounded
    interval \([0,\tmax]\) and that the density is
    bounded, so that we get the upper bound } &\qquad
                                                \le
                                                \int_{\mathcal{L}}
                                                \sum_{a=0,1}
                                                \bigg(
                                                \int_0^{\tmax}
                                                \big(                                                
                                                (p_{\lambda}-p_{\lambda_0})^2
                                                +
                                                \eta'^{-1}_0
                                                p_{\lambda}
                                                (S_0
                                                -
                                                S)
                                                \big)^2
                                                (t
                                                \mid
                                                a,
                                                \ell)
                                                dt
                                                \bigg)
                                                \pi^*(a
                                                \mid
                                                \ell)
                                                \mu_0(\ell)d\nu(\ell)
                                                ,
\end{align*}
so that
\( \Vert \lambda- \lambda_0 \Vert_{\pi^* \otimes \mu_0\otimes \rho} =
o_P(n^{-1/4})\) follows from
\( \Vert p - p_0 \Vert_{\pi^* \otimes \mu_0\otimes \rho} =
o_P(n^{-1/4})\).

\section*{Appendix D}

\renewcommand\theequation{D.\arabic{equation}}
\setcounter{equation}{0}

\subsection*{One-step targeted maximum likelihood for the target
  parameter}
\label{app:one:step:tmle}

The one-step TMLE \citep{van2016one,van2018targeted} uses a universal
least favorable submodel to solve the efficient influence curve
equation after only one update of the initial estimator.  Instead of
running the iterative TMLE as in Section \ref{sec:tmle}, we can use
the fluctuation model \eqref{eq:lambda:fluc} to generate a
corresponding universal loss function model.  The general recipe for
the construction of a universal least favorable submodel that solves
the efficient influence curve equation can be summarized largely as
follows. Say we have already defined a local least favorable submodel;
in our case, this is the fluctuation model defined in
\eqref{eq:lambda:fluc}. Now, instead of solving the score equation of
interest \eqref{eq:equation:written:out:0} (Section \ref{sec:tmle}),
we track the local least favorable submodel recursively with small
step \(d\eps\) until the desired influence curve equation is
solved. First it is checked which direction (\(\pm d\eps\)) that
decreases the value of the score equation. The first step then moves
\(d\eps\) in the relevant direction along the fluctuation model
\eqref{eq:lambda:fluc} evaluated in the initial estimator
\(\hat{\lambda}_n \). This gives \(\hat{\lambda}_n (\cdot; d\eps)\).
We then consider the fluctuation model through
\(\hat{\lambda}_n (\cdot; d\eps)\) and again move \(d\eps\) in the
relevant direction to obtain \(\hat{\lambda}_n (\cdot; 2d\eps)\). This
process is continued iteratively, at each step tracking the score
equation at zero fluctuation. We stop when we reach \(k^*\) such that
\(\hat{\lambda}_n (\cdot; k^*d\eps)\) solves the efficient influence
curve equation. The corresponding TMLE is obtained by plugging in
\(\hat{\lambda}_n (\cdot; k^*d\eps)\).

\end{document}